\documentclass[preprint,chaos,aps,amssymb,amsmath,floatfix]{revtex4}
\usepackage{graphicx}

\newcommand{\be}{\begin{equation}}
\newcommand{\ee}{\end{equation}}
\newcommand{\bi}{\bibitem}

\begin{document}

$\textheight=210mm 
$\textwidth=170mm

\begin{center}
Chaos 20 (2010) 023127
\end{center}

\title{Fractional Dissipative Standard Map}

\author{Vasily E. Tarasov}
\affiliation{Courant Institute of Mathematical Sciences,
New York University,
251 Mercer St., New York, NY 10012, USA\\
Skobeltsyn Institute of Nuclear Physics, 
Moscow State University, Moscow 119991, Russia 
}
\author{M. Edelman}
\affiliation{Courant Institute of Mathematical Sciences,
New York University,
251 Mercer St., New York, NY 10012, USA\\
Department of Physics, Stern College at Yeshiva University,
245 Lexington Ave, New York, NY 10016, USA
}

\begin{abstract}
Using kicked differential equations of motion with derivatives 
of noninteger orders, we obtain generalizations of the dissipative 
standard map. The main property of these generalized maps, 
which are called fractional maps, is long-term memory.
The memory effect in the fractional maps means
that their present state of evolution depends on all past states 
with special forms of weights. 
Already a small deviation of the order of derivative from the integer
value corresponding to the regular dissipative standard map 
(small memory effects) leads to the
qualitatively new behavior of the corresponding attractors.
The fractional dissipative standard maps are used to demonstrate a
new type of fractional attractors in the wide range of the fractional
orders of derivatives.

\end{abstract}

\maketitle

{\bf  
Discrete maps are widely used to study the general properties of
dynamical systems. In those cases when they can be derived from 
differential equations, their analysis gives the exact properties
of the corresponding systems. In this article we derive
discrete maps (fractional maps) from the fractional differential
equations, which correspond to the fractional generalizations of 
the dissipative standard map \cite{DM1}. We demonstrate how the attractors
of the fractional maps are different from the attractors 
of the dissipative standard map.
}

\section{Introduction}

There is a number of distinct areas of physics where
basic problems can be reduced to the study of simple discrete maps. 
Discrete maps  as substitutes of differential equations have been used 
to study evolution problems in Refs. \cite{Zbook1,Zbook3,Chirikov,Schuster,CE}. 
They lead to a simpler formalism, which is particularly useful in simulations.
The dissipative standard map is one of the most widely studied maps.
In this paper we consider fractional generalizations of the dissipative
standard map which are described by fractional differential equations 
\cite{SKM,Podlubny,KST}. 

The treatment of nonlinear dynamics in terms of discrete maps 
is a very important step in understanding the qualitative
behavior of systems described by differential equations.
The derivatives of noninteger orders are a natural generalization of 
the ordinary differentiation of the integer order. 
The fractional differentiation with respect to time is
characterized by long-term memory effects which 
correspond to intrinsic dissipative processes 
in physical systems. 
The application of memory effects to discrete maps means 
that their present state of evolution depends on all past states 
\cite{Ful,Fick1,Giona,Fick2,Gallas,Stan,JPA2008}. 

Discrete maps with memory can be derived (see Ref. \cite{JPA2008})
from equations of motion with fractional derivatives.
In Ref. \cite{JPA2008} a fractional generalization of the standard map 
has been derived from a fractional differential equation.
A fractional generalization of the
dissipative standard map was also suggested in \cite{JPA2008}.
Unfortunately, in that generalization a dissipation was introduced 
by the change of the variable $p_n \rightarrow -bp_n$.
The map equations were not directly connected 
with a fractional equation of motion.
In this paper we propose two generalizations  
of the dissipative standard map. The first one is derived from 
a differential equation with fractional damped kicks.
The second generalization of the dissipative standard map 
is derived from a fractional differential equation (kicks are not fractional) . 
A nonlinear system with fractional derivatives perturbed by a periodic force 
exhibits a new type 
of chaotic motion which can be called the fractional chaotic attractor \cite{ZSE}. 
Fractional discrete maps \cite{JPA2008} are used to study 
new types of attractors of fractional dynamics described by kicked 
fractional equations. 
In this paper some fractional differential equations of motion 
of kicked systems with friction are considered. 
Corresponding discrete maps with memory are derived from these equations. 
The fractional generalizations of the dissipative standard map are 
suggested and these maps are used in computer simulations.

\section{Discrete maps without memory}

In this section, a brief review of discrete maps 
is considered to fix notations and provide convenient references. 
For details, see Refs. \cite{Zbook1,Zbook3,Chirikov,Schuster,CE}.

\subsection{Standard map}

Let us consider the equation of motion
\be \label{eq1}
\ddot{x} + K \sin(x) \sum^{\infty}_{n=0} \delta \Bigl(t-n \Bigr)=0,
\ee
in which perturbation is a periodic sequence of delta-function-type pulses (kicks)
following with period $T=1$, $K$ is the amplitude of the pulses.
This equation can be presented in the Hamiltonian form
\be \label{HE1}
\dot{x}=p , \quad
\dot{p}+K \sin(x) \sum^{\infty}_{n=0} \delta \Bigl(t-n \Bigr)=0 .
\ee
It is well-known that these equations can be represented 
(see, for example, Chapter 5 in Ref. \cite{Zbook3}) in the form of discrete map
\be \label{E1}
x_{n+1}=x_n+p_{n+1} , \ee
\be \label{E2}
p_{n+1}=p_n-K\, \sin(x_n) .
\ee
Equations (\ref{E1}) and (\ref{E2}) are called the standard map.
This map is also called the Chirikov map \cite{Chirikov}.


\subsection{Dissipative standard map}

The dissipative standard map \cite{DM1,DM2,DM3} is 
\be \label{Zasl1a}
X_{n+1} = X_n +  \mu Y_{n+1}+\Omega , 
\ee
\be \label{Zasl2a}
Y_{n+1} = e^{-q} \, [ Y_n + \varepsilon \, \sin(X_n) ] ,
\ee
where $ \mu=(e^q-1)/q. $
The dissipative standard map is also called the Zaslavsky map.
Note that a shift $\Omega$ does not play an important role 
and it can be put to zero ($\Omega=0$). 
The dissipative standard map with $\Omega=0$ can be represented by the equations
\be \label{Zasl1}
X_{n+1}=X_n+P_{n+1} ,
\ee
\be \label{Zasl2}
P_{n+1}=-bP_n-Z \, \sin(X_n) .
\ee
For the parameters
\be \label{cond} Z= - \varepsilon  \mu e^{-q} , \quad 
P_n= \mu Y_n , \quad b=-e^{-q}
\ee
equations (\ref{Zasl1}) and (\ref{Zasl2}) give
Eqs. (\ref{Zasl1a}) and (\ref{Zasl2a}) with $\Omega=0$. 

For $b=-1$ and $Z=K$, we get the standard map 
which is described by Eqs. (\ref{E1}) and (\ref{E2}) with $T=1$.

Note that for large $q \rightarrow \infty$ (for small $b \rightarrow 0$)
Eqs. (\ref{Zasl1}) and (\ref{Zasl2}) with $Z=-K$
shrink to the one-dimensional sine map proposed by Arnold \cite{Ar}, 
\be \label{sine-map} 
X_{n+1}=X_{n}+K \sin (X_n).
\ee

\subsection{Kicked damped rotator map}

The equation of motion for a kicked damped rotator is
\be \label{dr-0}
\ddot{x}+ q \dot{x}=K G(x) \sum^{\infty}_{n=0} \delta \Bigl(t -n T \Bigr) .
\ee
It is well known \cite{Schuster} that Eq. (\ref{dr-0}) gives 
the two-dimensional map
\be \label{dr-map1}
x_{n+1}=x_n+\frac{1-e^{-qT}}{q} [ p_n+K G(x_n)] ,
\ee
\be \label{dr-map2}
p_{n+1}=e^{-qT} [ p_n+K G(x_n)] .
\ee
This map is known as the kicked damped rotator map. 
The phase volume shrinks each time step by a factor $\exp(-q)$. 
The map is defined by two important parameters, 
dissipation constant $q$ and force amplitude $K$. 
These equations can be rewritten in the form
\[ x_{n+1}=x_n+\frac{e^{qT}-1}{q} p_{n+1} , \]
\[ p_{n+1}=e^{-qT} [ p_n+K G(x_n)] . \]
It is easy to see that these equations give
the dissipative standard map (\ref{Zasl1}) and (\ref{Zasl2}) with $\Omega=0$ 
if we use
\[ X_n=x_n, \quad Y_n=p_n, \quad \varepsilon=K, \quad T=1, \quad G(x)=\sin(x). \]
This allows us to derive dissipative standard map (\ref{Zasl1a}) and (\ref{Zasl2a}) 
from the differential equation,
\be \label{dr-0b}
\ddot{X}+ q \dot{X}=\varepsilon \sin(X) 
\sum^{\infty}_{n=0} \delta \Bigl(t -n \Bigr) .
\ee
These equations give the discrete map 
defined by Eqs. (\ref{Zasl1a}) and (\ref{Zasl2a}) with $\Omega=0$.


\section{Fractional standard map and dissipation}

A fractional generalization of 
the differential equation (\ref{eq1}) has been suggested in Ref. \cite{JPA2008}. 
The discrete map which corresponds to the fractional equation 
of order $1 <\alpha \le 2$ was derived. 
This map can be considered as a generalization of the standard map
for the case $1 <\alpha \le 2$.

We consider a fractional generalization of Eq. (\ref{eq1}) in the form
\be \label{eq2}
_0D^{\alpha}_t x+K \sin(x) \sum^{\infty}_{n=0} \delta \Bigl(t-n \Bigr)=0, 
\quad (1 <\alpha \le 2) ,
\ee
where $ _0D^{\alpha}_t$ is the Riemann-Liouville fractional derivative
\cite{SKM,Podlubny,KST}, which is defined by
\be \label{RLFD}
_0D^{\alpha}_t x=D^2_t \ _0I^{2-\alpha}_t x=
\frac{1}{\Gamma(2-\alpha)} \frac{d^2}{dt^2} \int^{t}_0 
\frac{x(\tau) d \tau}{(t-\tau)^{\alpha-1}} , \quad (1 <\alpha \le 2) .
\ee
Here we use the notation $D^2_t=d^2/dt^2$, and 
$ _0I^{\alpha}_t$ is a fractional integration \cite{SKM,Podlubny,KST}. 

Defining the momentum as
\[ p(t)= \, _0D^{\alpha-1}_t x(t), \]
and using the initial conditions
\be 
(_0D^{\alpha-1}_tx) (0+) = p_1, \quad (_0D^{\alpha-2}_tx) (0+) = b,
\ee
it is possible to derive the equation for the fractional standard map. \\

\noindent
{\large \bf Proposition 1.}
{\it The fractional differential equation of the kicked system (\ref{eq2})
is equivalent to the discrete map
\be \label{N1}
x_{n+1}=\frac{1}{\Gamma(\alpha)} \sum^{n}_{k=0} p_{k+1} V_{\alpha}(n-k+1) 
+\frac{b}{\Gamma(\alpha-1)} (n+1)^{\alpha-2} ,
\ee
\be 
p_{n+1}=p_n- \, K \sin(x_n) , \quad (1<\alpha\le 2) ,
\ee
where the function $V_{\alpha}(z)$ is defined by}
\be 
V_{\alpha}(z)=z^{\alpha-1}-(z-1)^{\alpha-1} .
\ee

Proof of this Proposition is given in Ref. \cite{TarV}. \\

A fractional generalization of the dissipative standard map 
suggested in Refs. \cite{DM1,DM2} 
can be defined by 
\be \label{Zaslmap1}
x_{n+1}=  \frac{1}{\Gamma(\alpha)} 
\sum^{n}_{k=0} p_{k+1} V_{\alpha}(n-k+1), \quad (1<\alpha\le 2),
\ee
\be \label{Zaslmap2}
p_{n+1}= -bp_n-Z \, \sin ( x_n ) ,
\ee
where the parameters are defined by conditions (\ref{cond}). 
For $b=-1$ and $Z=K$ Eqs. (\ref{Zaslmap1}) and (\ref{Zaslmap2}) give 
the fractional standard map with $T=1$. 
Note that this fractional dissipative standard map is not derived from 
a fractional differential equation. 
This map is derived by $p_n \rightarrow -bp_n$ in 
the fractional standard map. 
Fractional dissipative standard map can be derived 
from fractional differential equations.
In this paper, we derive two fractional generalizations 
of the dissipative standard map which are obtained from fractional
differential equations.


\section{Fractional derivative in the kicked term 
and the first fractional dissipative standard map}

In this section we suggest the first fractional generalization of 
differential equation (\ref{dr-0}) for a kicked damped rotator. 
In this generalization we introduce a fractional derivative in 
the kicked damped term, i.e. the term 
of a periodic sequence of delta-function-type pulses (kicks),
and derive the corresponding discrete map.

Consider the fractional generalization of equation (\ref{dr-0}) in the form 
\be \label{eq7-n}
D^2_t X(t) - q D^1_t X(t)=
\varepsilon \sin \left(\, _0^CD^{\beta}_t X \right) \sum^{\infty}_{n=0} \delta (t-n), 
\quad (0 \le \beta <1 ) ,
\ee
where $ q \in \mathbb{R}$, 
and $ _0^CD^{\beta}_t$ is the Caputo fractional derivative 
\cite{KST} of the order $0 \le \beta<1$ defined by 
\be \label{CFD}
_0^CD^{\beta}_t X=\ _0I^{1-\beta}_t D^1_t X=
\frac{1}{\Gamma(1-\beta)} \int^{t}_0 
\frac{d \tau}{(t-\tau)^{\beta}} \frac{d X(\tau)}{d \tau} ,
\quad (0 \le \beta<1).
\ee
Here we use the notation $D^1_t=d/dt$, and 
$ _0I^{\alpha}_t$ is a fractional integration \cite{SKM,Podlubny,KST}. 
For $\beta=0$ fractional equation (\ref{eq7-n}) gives equation (\ref{dr-0}). 
Note that we use the minus on the left-hand side of Eq. (\ref{eq7-n}), 
where $q$ can be a positive or negative value. 
Fractional derivative $ _0^CD^{\beta}_t X$ 
is presented in the kicked damped term. \\

\noindent
{\large \bf Proposition 2.}
{\it The fractional differential equation of the kicked system (\ref{eq7-n})
is equivalent to the discrete map
\be \label{FDR3-n}
X_{n+1}=X_n+\frac{1-e^{-q}}{q} Y_{n+1} ,
\ee
\be \label{FDR4-n}
Y_{n+1}=e^{q} \, \Bigl[ Y_n+\varepsilon 
\sin \Bigl(\frac{1}{\Gamma(1-\beta)} 
\sum^{n-1}_{k=0} \, Y_{k+1} \, W_{2-\beta}(q,k-n) \Bigr) \Bigr] ,
\ee
where the functions $W_{2-\beta}(a,b)$ are defined by 
\[ W_{2-\beta}(a,b)=a^{\beta-1}  e^{a(b+1)} 
\Bigl[ \Gamma(1-\beta, ab ) - \Gamma(1-\beta, a(b+1) ) \Bigr] , \]
and $\Gamma(a,b)$ is the incomplete Gamma function }
\be \label{incomplete}
\Gamma(a,b)=\int^{\infty}_b y^{a-1} e^{-y} dy .
\ee

{\bf Proof.} 
Fractional equation (\ref{eq7-n}) can be presented in the Hamiltonian form
\[ \dot{X}=Y , \]
\be \label{HE3-n}
\dot{Y}-q Y=
\varepsilon \sin(\, _0^CD^{\beta}_t X ) \sum^{\infty}_{n=0} \delta (t-n), 
\ee
where $0 <\beta< 0$, and $q \in \mathbb{R}$.

Between any two kicks, 
\be \label{67-n}
\dot{Y}-q Y=0 . 
\ee
For $t\in(t_n+0,t_{n+1}-0)$, the solution of Eq. (\ref{67-n}) is
\be \label{eq4-n}
Y(t_{n+1}-0)=Y(t_n+0) e^{q} .
\ee
Let us use the notations $t_n=nT$, with $T=1$ and
\[ X_n=X(t_n-0)=\lim_{\epsilon \rightarrow 0} X(n-\epsilon), \]
\be \label{not3-n}
Y_n=Y(t_n-0)=\lim_{\epsilon \rightarrow 0} Y(n-\epsilon) .
\ee
For $t\in (t_n-\epsilon,t_{n+1}-\epsilon)$, 
the general solution of Eq. (\ref{HE3-n}) is
\be
Y(t)=Y_n e^{q(t-t_n)}+ 
\varepsilon \sum^{\infty}_{m=0} \sin (\, _0^CD^{\beta}_{t_m} X )  
\int^t_{t_n-\epsilon} d \tau e^{q(t-\tau)} \delta (\tau-m) .
\ee
Then
\be \label{eq5-n}
Y_{n+1}= e^{q} \Bigl[ Y_n+ \varepsilon \sin(\, _0^CD^{\beta}_{t_n} X ) \Bigr] .
\ee
Using Eq. (\ref{eq5-n}), the integration of the first equation of (\ref{HE3-n}) gives
\be \label{eq6-n}
X_{n+1}=X_n - \frac{1-e^{q}}{q} 
\Bigl[ Y_n+ \varepsilon \sin(\, _0^CD^{\beta}_{t_n} X ) \Bigr] .
\ee
Let us consider the Caputo fractional derivative 
from Eqs. (\ref{eq5-n}) and (\ref{eq6-n}). 
It is defined by the equation
\[ _0^CD^{\beta}_{t_n} X= _0I^{1-\beta}_t D^1_t X=
\frac{1}{\Gamma(1-\beta)} \int^{t_n}_0 
\frac{d \tau}{(t_n-\tau)^{\beta}} \frac{d X(\tau)}{d \tau} , 
\quad (0\le \beta <1). \]
Using $Y(\tau)= dX(\tau)/d\tau$, this relation can be rewritten as
\be \label{tk3-n}
_0^CD^{\beta}_{t_n} X= \frac{1}{\Gamma(1-\beta)} 
\sum^{n-1}_{k=0} \int^{t_{k+1}}_{t_k} 
\frac{ Y(\tau) d \tau}{(t_n-\tau)^{\beta}} ,
\ee
where $t_{k+1}=t_k+1=(k+1)$, and $t_k=k$, such that $t_0=0$.
For $\tau \in (t_k,t_{k+1})$, equations (\ref{eq4-n}) and (\ref{not3-n}) give
\[ Y(\tau)=Y(t_k+0) e^{q(\tau-t_k)}=
Y(t_{k+1}-0) e^{-q} e^{q(\tau-t_k)}= \]
\[ =Y_{k+1} e^{q(\tau-t_k-1)}=
Y_{k+1} e^{q(\tau-t_{k+1})} . \]
Then 
\[ 
\int^{t_{k+1}}_{t_k} \frac{ Y(\tau) d \tau}{(t_n-\tau)^{\beta}} =
Y_{k+1} \int^{t_{k+1}}_{t_k} 
e^{q(\tau-t_{k+1})} (t_n-\tau)^{-\beta} d\tau =
\]

\[ 
=Y_{k+1} \int^{t_n-t_k}_{t_n-t_{k+1}} 
e^{q(t_n-t_{k+1}-z)} z^{-\beta} dz =
Y_{k+1} e^{q(t_n-t_{k+1})}
\int^{t_n-t_k}_{t_n-t_{k+1}} z^{-\beta} e^{-qz} dz =
\]

\be \label{y-n}
= Y_{k+1} q^{\beta-1}  e^{q(n-k-1)}
\int^{q(t_n-t_k)}_{q(t_n-t_{k+1})} y^{-\beta} e^{-y} dy .
\ee
As a result, equation (\ref{y-n}) gives
\[ \int^{t_{k+1}}_{t_k} \frac{ Y(\tau) d \tau}{(t_n-\tau)^{\beta}} = \]
\be \label{p11-n}
=Y_{k+1} q^{\beta-1} e^{q(n-k-1)}
\Bigl[ \Gamma(1-\beta, q(t_n-t_{k+1}) ) - \Gamma(1-\beta, q(t_n-t_k) ) \Bigr] .
\ee
Here $\Gamma(a,b)$ is the incomplete Gamma function (\ref{incomplete}), 
where $a$ and $b$ are complex numbers. 
Using (\ref{tk3-n}) and (\ref{p11-n}), we obtain
\be \label{tk4-n}
_0^CD^{\beta}_{t_n} X= 
\frac{1}{\Gamma(1-\beta)} \sum^{n-1}_{k=0} Y_{k+1} W_{2-\beta}(q,k-n) ,
\quad (0 \le \beta <1) ,
\ee
where
\be \label{Wa-n}
W_{2-\beta}(a,b)=a^{\beta-1} e^{a(b+1)} 
\Bigl[ \Gamma(1-\beta, ab ) - \Gamma(1-\beta, a(b+1) ) \Bigr] .
\ee
Substitution of Eq. (\ref{tk4-n}) into Eqs. (\ref{eq5-n}) and (\ref{eq6-n}) gives
\be \label{FDR1-n}
Y_{n+1}=e^{q} \Bigl[ Y_n+\varepsilon
\sin \Bigl( \frac{1}{\Gamma(1-\beta)} 
\sum^{n-1}_{k=0} Y_{k+1} W_{2-\beta}(q,k-n) \Bigr) \Bigr] ,
\ee
\be \label{FDR2-n}
X_{n+1}=X_n-\frac{1-e^{q}}{q} \Bigl[Y_n+
\varepsilon \sin \Bigl(\frac{1}{\Gamma(1-\beta)} 
\sum^{n-1}_{k=0} Y_{k+1} W_{2-\beta}(q,k-n) \Bigr) \Bigr] .
\ee
Equations (\ref{FDR1-n}) and (\ref{FDR2-n}) can be presented  
in the form of Eqs. (\ref{FDR3-n}) and (\ref{FDR4-n}).

This ends the proof. 
$\ \ \ \Box$ \\

The iteration equations (\ref{FDR3-n}) and (\ref{FDR4-n})
define a fractional generalization of  the dissipative standard map.
For $\beta=0$ this map gives the Zaslavsky map 
(\ref{Zasl1a}) and (\ref{Zasl2a}) with 
\be
\mu = (1-e^{-q})/q 
\ee
and $\Omega=0$.


\section{Fractional derivative in the unkicked terms 
and the second fractional dissipative standard map}

In this section we suggest a fractional generalization of 
the differential equation for a kicked damped rotator
with fractional derivatives 
in the unkicked terms and derive the
corresponding discrete map.

We consider the fractional generalization of equation (\ref{dr-0}) in the form 
\be \label{fdr}
_0D^{\alpha}_t X(t) - q \, _0D^{\beta}_t X(t) =
\varepsilon \sin(X) \sum^{\infty}_{n=0} \delta (t -n ) ,
\ee
where
\[ q \in \mathbb{R}, \quad 1< \alpha\le 2 , \quad \beta=\alpha-1 , \]
and $ _0D^{\alpha}_t$ is the Riemann-Liouville fractional derivative
\cite{SKM,Podlubny,KST}, which is defined by Eq. (\ref{RLFD}).
This equation has fractional derivatives in the unkicked terms, i.e. 
on the left-hand side of Eq. (\ref{fdr}). 
We use the minus in the left-hand side of Eq. (\ref{fdr}), 
where $q$ can have a positive or negative value. \\

\noindent
{\large \bf Proposition 3.}
{\it The fractional differential equation of the kicked system (\ref{fdr})
is equivalent to the discrete map
\be \label{FDM2-1}
X_{n+1}=\frac{1}{\Gamma(\alpha-1)} 
\sum^{n}_{k=0} \, Y_{k+1} \, W_{\alpha}(q,k-n-1) ,
\ee
\be \label{FDM2-2}
Y_{n+1}=e^{q} \, \Bigl[ Y_n+ \varepsilon \sin ( X_n ) \Bigr] ,
\ee
where the functions $W_{\alpha}(a,b)$ are defined by 
\be \label{Wa}
W_{\alpha}(a,b)=a^{1-\alpha}  e^{a(b+1)} \,
\Bigl[ \Gamma(\alpha-1, ab ) - \Gamma(\alpha-1, a(b+1) ) \Bigr] ,
\ee
and $\Gamma(a,b)$ is the incomplete Gamma function (\ref{incomplete}).} \\

{\bf Proof.}
Let us define an auxiliary variable $\xi (t)$ such that
\be \label{xi-2}
_0^CD^{2-\alpha}_t \xi=X(t),
\ee
where $ _0^CD^{2-\alpha}_t$ is the Caputo fractional derivative (\ref{CFD}). 
Using 
\be
_0I^{2-\alpha}_t\ _0^CD^{2-\alpha}_t \xi = \xi(t)-\xi(0) , \quad (0\le 2-\alpha <1) ,
\ee
we obtain
\be \label{xix-2}
_0D^{\alpha}_t X= D^2_t \ _0I^{2-\alpha}_t X=
D^2_t \ _0I^{2-\alpha}_t \ _0^CD^{2-\alpha}_t \xi=
D^2_t ( \xi(t)-\xi(0) )=D^2_t \xi ,
\ee
and 
\[
_0D^{\beta}_t X= D^1_t \ _0I^{1-\beta}_t X=
D^1_t \ _0I^{2-\alpha}_t X=
\]
\be \label{xix-3}
=D^1_t \ _0I^{2-\alpha}_t \ _0^CD^{2-\alpha}_t \xi
=D^1_t ( \xi(t)-\xi(0) )=D^1_t \xi .
\ee
Substitution of Eqs. (\ref{xix-2}), (\ref{xix-3}), and (\ref{xi-2}) 
into Eq. (\ref{fdr}) gives
\be \label{eq7}
D^2_t \xi - q D^1_t \xi=
\varepsilon \sin(\, _0^CD^{2-\alpha}_t \xi ) \sum^{\infty}_{n=0} \delta (t-n), 
\quad (1 <\alpha \le 2) .
\ee
The fractional equation (\ref{eq7}) can be presented in the Hamiltonian form
\[
\dot{\xi}=Y ,
\]
\be \label{HE3}
\dot{Y}-q Y=
\varepsilon \sin (\, _0^CD^{2-\alpha}_t \xi ) \sum^{\infty}_{n=0} \delta (t-n), 
\quad (1 <\alpha < 2, \quad q \in \mathbb{R}) .
\ee
Using Eq. (\ref{FDR4-n}) of Proposition 2, we obtain
\[ Y_{n+1}=e^{q} \Bigl[ Y_n + \varepsilon \sin \Bigl(\frac{1}{\Gamma(\alpha-1)} 
\sum^{n-1}_{k=0} Y_{k+1} W_{\alpha}(q,k-n) \Bigr) \Bigr] . \]
For $(X_n,Y_n)$, we use equation (\ref{tk4-n}) in the form
\[ X_n=\, _0^CD^{2-\alpha}_{t_n} \xi
=\frac{1}{\Gamma(\alpha-1)} \sum^{n-1}_{k=0} Y_{k+1} W_{\alpha}(q,k-n)  . \]
As a result, we have
\be \label{New1}
X_{n+1}=\frac{1}{\Gamma(\alpha-1)} 
\sum^{n}_{k=0} Y_{k+1} W_{\alpha}(q,k-n-1) ,
\ee
\be \label{New2}
Y_{n+1}=e^{q} \Bigl[ Y_n+ \varepsilon \sin ( X_n ) \Bigr] ,
\ee
where $W_{\alpha}(a,b)$ is defined in Eq. (\ref{Wa}).
This ends the proof. 
$\ \ \ \Box$ \\

\begin{figure}
\centering
\rotatebox{0}{\includegraphics[width=12 cm]{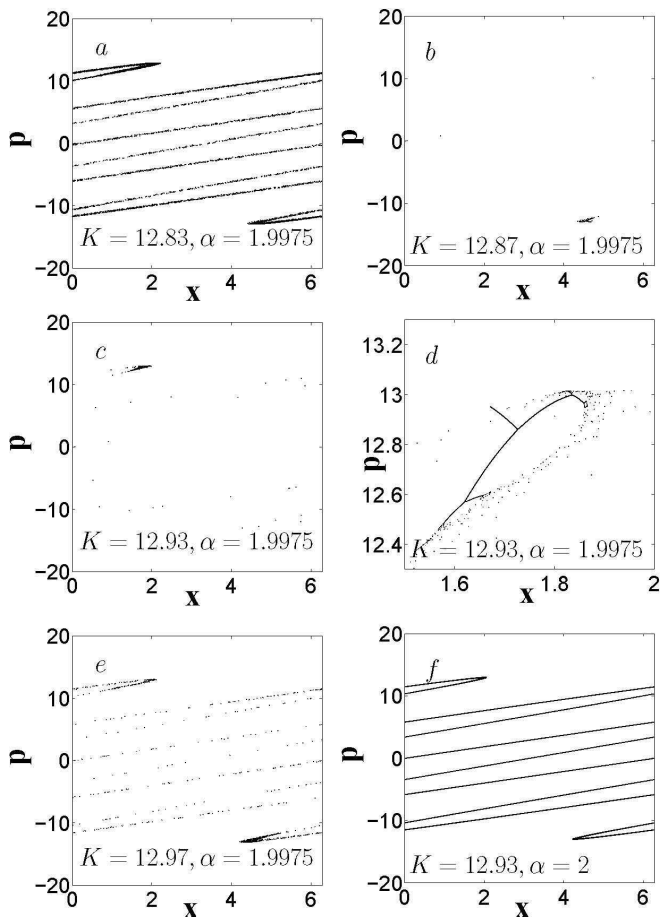}}
\caption{\label{Fig1} Structures of the chaotic attractors for different values of $K$ 
obtained after $10^5$ iterations ($\Gamma = 5$, $\Omega = 0$); $\alpha=1.9975$ 
in (a)-(e). 
(a) $K=12.83$; (b) $K=12.87$; (c) $K=12.93$; (d) zoom of (c); 
(e) $K=12.97$; (f) $\alpha=2$, $K=12.93$.}
\end{figure}

\begin{figure}
\centering
\rotatebox{0}{\includegraphics[width=12 cm]{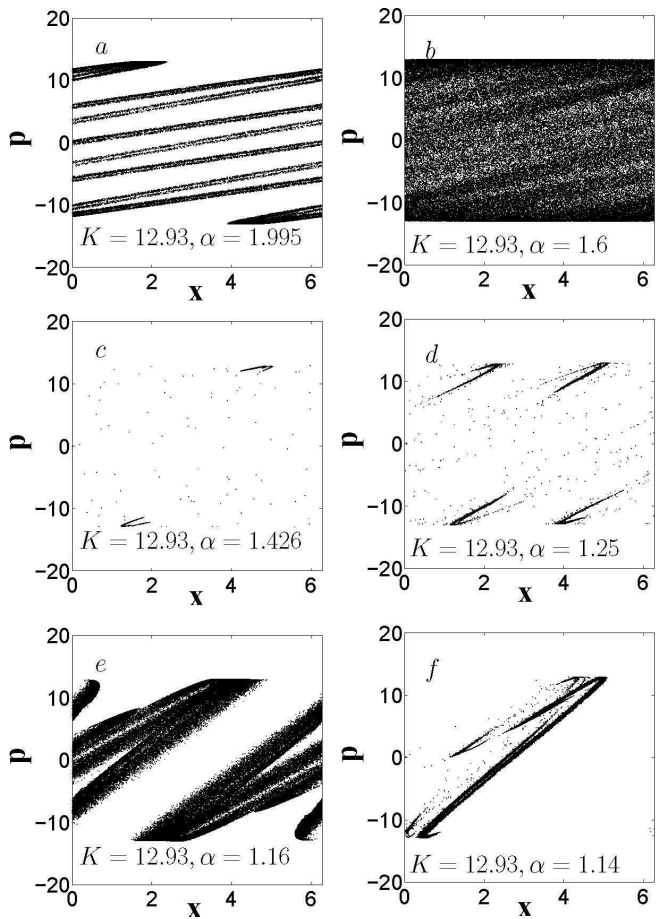}}
\caption{\label{Fig2} Structures of the fractional chaotic attractors for $K=12.93$ 
and different values of $\alpha$ obtained after $10^5$ 
iterations ($\Gamma = 5$, $\Omega = 0$).}
\end{figure}


If we use the variables
\[ P_n= \mu Y_n , \quad b = - e^q , \quad Z = - \mu \varepsilon e^q , \]
then equations (\ref{FDM2-1}) and (\ref{FDM2-2}) give  
\be  \label{T1}
X_{n+1} = \frac{\mu^{-1}}{\Gamma(\alpha-1)} 
\sum^{n}_{k=0} P_{k+1} W_{\alpha}(q,k-n-1) .
\ee
\be  \label{T2}
P_{n+1}= - b P_n -Z \sin ( X_n ) ,
\ee
These equations can be considered as a fractional generalization 
of the dissipative standard map 
equations (\ref{Zasl1}) and (\ref{Zasl2}) with $\Omega=0$.
For $\alpha=2$, this fractional dissipative standard  map gives 
the dissipative standard map that 
is described by Eqs. (\ref{Zasl1}) and (\ref{Zasl2}).

\section{Numerical simulations}

Numerical simulations were performed for the second fractional 
dissipative standard map (Eqs.  (\ref{T1}) and (\ref{T2})). 
First we used our code to reproduce the results presented in 
Fig.~1 from \cite{DM3} for the 
structures of the chaotic attractors of the dissipative standard 
map at the window of the ballistic motion near 
$K \approx 4\pi$ ($q=-5$, $K=\varepsilon \exp(q)$, 
and used in Ref. \cite{DM3} $\Gamma$ is equal to $-q$) 
for the fractional standard map with
$\alpha = 2$ and obtained a perfect agreement 
(an example is given in Fig.~1(f)). 
As $\alpha$ decreases slightly from $\alpha=2$ to $\alpha=1.9975$, 
the window of the ballistic motion 
shrinks and moves to the higher values of $K$. 
Already for $\alpha=1.9975$ in Figs.~1(a)-1(e) 
the ballistic motion appears for $K>12.86$ and disappears at $K=12.97$. 
The window is completely closed at $\alpha \approx 1.9969$. 
The structures of two symmetric attractors with disjoint basins 
which appear within the window (Figs.~1(b) and 1(c)) is also very different 
from the structures of the dying attractors of the  dissipative standard 
map \cite{DM1,DM3}. 
The attractor in Fig.~1(d) evolves from period 8 trajectory to period 4, 
period 2, and, finally, period 1 trajectory slowly moving 
in the direction of the upper left corner 
with the step of the order of $10^{-7}$.

\begin{figure}
\centering
\rotatebox{0}{\includegraphics[width=12 cm]{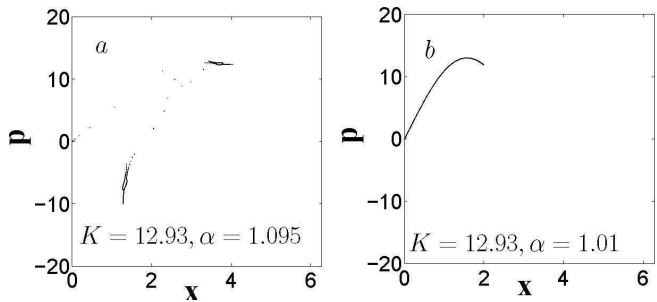}}
\caption{\label{Fig3} Attracting trajectories for $K=12.93$ and small 
values of $\alpha$ obtained after $10^5$ iterations ($\Gamma = 5$, $\Omega = 0$).}
\end{figure}

When $\alpha$ decreases further, the structures of the fractional chaotic 
attractors evolve in the manner presented in Fig.~2, 
where one can find one-scroll, two-scroll, and four-scroll 
fractional chaotic attractors, strongly deviating from the chaotic attractor 
of the dissipative standard map Fig.~1(f) (see also \cite{DM3}). 
The problem of existence of multiscroll fractional chaotic attractors 
was considered in Ref. \cite{TH} but for the fractional differential
equations with the Caputo derivatives.
For values of $\alpha$ near 1 fractional chaotic attractor turns into 
period 2 and for smaller values period 1 attracting trajectories, Fig.~3.

\section{Conclusion}

The suggested discrete maps with memory are generalizations 
of the dissipative standard map.
These maps are potentially useful for description of 
fractional dynamics of complex physical systems. 
We consider viscoelastic and dielectric materials 
as examples of complex media (physical systems) 
whose dynamics could be described by 
the fractional dissipative standard maps and 
corresponding fractional differential equations. 
This assumption is based on the following.

(1) It is well known that viscoelastic materials
can be described by fractional differential equations
(see, for example, Ref. \cite{MBook}).
The fractional dissipative standard map could be employed to model 
the one-dimensional simplification of the equations of viscoelastic materials 
in which a perturbation is a periodic sequence of delta-function-type
pulses (kicks) following with some period.

(2) For a wide class of dielectric materials
the dielectric susceptibility follows
a fractional power-law frequency dependence, which is called
the "universal" response \cite{Jo2,Jrev} over extended frequency ranges.
The electromagnetic fields in such dielectric media
are described by differential equations
with fractional time derivatives \cite{JPCM2008-1,JPCM2008-2}.
These fractional equations for electromagnetic waves
in dielectric media are common to a wide class of materials,
regardless of the type of physical structure,
or of the nature of the polarizing species.
We assume that the fractional maps could be applied to dielectric media
in which a perturbation is a periodic sequence of kicks.

The suggested fractional dissipative standard maps demonstrate 
a chaotic behavior with a new type of attractors.
The interesting property of these fractional maps is long-term memory.
As a result, a present state of evolution depends on all 
past states with the weight functions.
The fractional dissipative standard maps are equivalent 
to the correspondent fractional kicked differential equations.
Note that to derive discrete maps an approximation for 
fractional derivatives of these equations is not used.

Computer simulations of the suggested discrete maps 
with memory prove that the nonlinear dynamical systems,
which are described by the equations with fractional 
derivatives, exhibit a new type of chaotic motion. 
This type of motion demonstrates a fractional generalization of attractors.

It has been shown in Ref. \cite{PLA2009} that in the case 
$q=0$ (fractional standard map) the fractional discrete map
demonstrates a new type of attractors such as
slow converging and slow diverging trajectories, ballistic trajectories, 
fractal-like structures, and chaotic trajectories.
At least one type of fractal-like sticky attractors
in the chaotic sea can be observed \cite{PLA2009}
for the fractional standard map.
The properties of stability and existence of the 
fractional attractors in the asymptotic sense for $q=0$ have also been
described in Ref. \cite{PLA2009}. The attractors presented in 
Figs.~1-3 are quite different from the corresponding 
regular chaotic attractors and $q=0$ attractors.
The detail classification of these attractors and corresponding chaotic motion
will be considered in the nearest future and
published in the next paper.


\section*{Acknowledgments}

We express our gratitude to H. Weitzner for many comments and helpful discussions.
This work was supported by the Office of Naval Research, 
Grant No. N00014-02-1-0056.



\end{document}